\documentclass
[aps,prl,twocolumn,floatfix,english,showpacs,10pt,superscriptaddress]{revtex4-2}
\usepackage{graphicx}
\usepackage{booktabs}
\usepackage{amsmath}
\usepackage{physics}
\usepackage{amssymb}
\usepackage{colordvi}
\usepackage{verbatim}
\usepackage{xcolor}
\usepackage{mathrsfs}
\usepackage{epsfig}
\usepackage{lipsum}
\usepackage{amsfonts}
\usepackage{makecell}
\usepackage{esint}
\usepackage[unicode=true, breaklinks=false, pdfborder={0 0 1}, backref=false,
colorlinks=true, linkcolor=blue, urlcolor=blue, citecolor=blue]{hyperref}%
\providecommand{\U}[1]{\protect\rule{.1in}{.1in}}
\setcitestyle{numbers,square}
\begin{document}

\title{Angular-Momentum Swapping Mediated by Spin-Orbital Textures in Centrosymmetric 
Multiorbital Materials}

\author{Chengyuan Cai}
\thanks{These authors contributed equally to this work.}
\affiliation{School of Physics, Huazhong University of Science and Technology, Wuhan 430074, China}

\author{Hanchen Wang}
\thanks{These authors contributed equally to this work.}
\affiliation{Department of Materials, ETH Zurich, Zurich 8093, Switzerland}

\author{Rui-Chun Xiao}
\affiliation{Institute of Physical Science and Information Technology, Anhui University, Hefei 230601, China}
\affiliation{Anhui Provincial Key Laboratory of Magnetic Functional Materials and Devices, Anhui University, Hefei 230601, China}

\author{Tao Yu}
\email{taoyuphy@hust.edu.cn}
\affiliation{School of Physics, Huazhong University of Science and Technology, Wuhan 430074, China}

\date{\today}

\begin{abstract}
Spin swapping, which interchanges the directions of spin-current flow and spin polarization, provides an important route to engineering angular-momentum transport. Yet every established route relies on either strong impurity spin-orbit scattering or spin-band splitting. Here we uncover a fundamentally distinct route to spin swapping in magnetization-driven centrosymmetric multiorbital materials \textit{without any spin-band splitting}, where the orbital degree of freedom plays an unexpected role.
When the dynamical magnetization pumps a longitudinal primary spin current flowing normal to the interface, spin-orbit-entangled orbital textures redirect the injected spins into longitudinal and transverse orbital currents. Their subsequent conversion back to the spin sector generates transverse spin currents with propagation and polarization directions swapped relative to the primary spin current.
The swapping efficiency achieves an unprecedented $50\%$ for Ge. Governed by a fourth-rank tensor constrained by crystalline symmetry and without requiring spin-orbit scattering or inversion symmetry breaking, our finding reveals spin-orbital textures as hidden intermediate channels of angular-momentum swapping in a large class of unexplored materials.
\end{abstract}

\maketitle

\textit{Introduction}.---Angular-momentum currents with diverse spin or orbital polarizations are useful in information processing and memory~\cite{Hirsch1999,Sinova2015,Manchon2019,Go2021EPL,AtenciaReview2024,JoReview2024,Ando2025,Fukami2026}, a functionality achieved by so-called spin swapping, which converts a primary spin current into a secondary one by interchanging the directions of spin flow and spin polarization~\cite{Lifshits2009}.
Originally proposed as an extrinsic effect arising from spin-orbital impurity scattering~\cite{Lifshits2009,Shen2015}, it has been observed in Rashba-interface heterostructures and ferromagnet/nonmagnet bilayers~\cite{Kim2020,Fu2022,Ryu2022,Hazra2023,Wang2024,Kim2025,Wu2026} and harnessed to generate nontrivial (out-of-plane) spin polarizations for spin-orbital torque and anisotropic information transport. Spin-orbital scattering can be strong in disordered conductors~\cite{Saidaoui2015,Kokkeler2025}, superconductors~\cite{Espedal2017,Huang2018,Kamra2024}, magnetic bilayers~\cite{Saidaoui2016}, and ferromagnets~\cite{Ortiz2018,Cui2023}. Spin swapping can also arise intrinsically from band spin splitting by spin-orbit coupling (SOC) in bulk and at interfaces~\cite{Sadjina2012,Borge2017,Li2019}. Recently, spin swapping in ferromagnetic metals was predicted to be mediated by orbital textures~\cite{Park2022}, but in this mechanism the band exchange spin splitting is essential. It is then clear that spin swapping is difficult to realize in centrosymmetric non-magnetic materials.

\begin{figure}[htp!]
\centering
\includegraphics[width=0.45\textwidth]{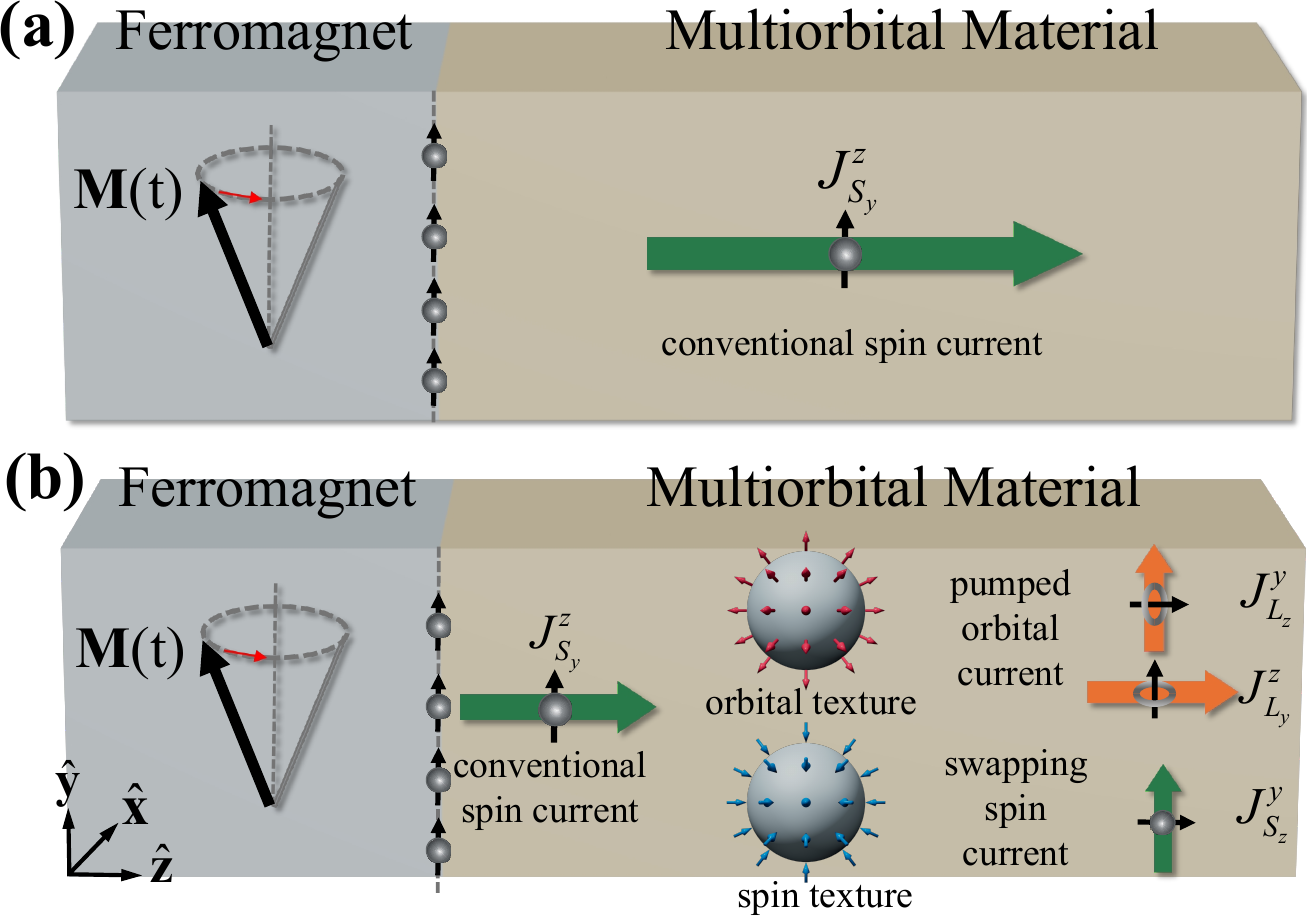}
\caption{Spin and orbital swapping in multiorbital materials pumped by magnetization ${\bf M}(t)$. (a): Pumped conventional spin current $J_{S_y}^{z}$ into conductors, polarized along the saturation magnetization $\hat{\bf y}$ and flowing normal to the interface. (b): The intrinsic orbital texture induces the spin texture by SOC, giving rise to a spin-swapping current $J_{S_z}^{y}$, polarized along $\hat{\bf z}$ and flowing along $\hat{\bf y}$. Meanwhile, longitudinal and transverse orbital currents $J_{L_y}^{z}$ and $J_{L_z}^{y}$ are generated.}
\label{spin_injection}
\end{figure}

Here we overturn this view: in the pumping configuration (Fig.~\ref{spin_injection}), spin pumping~\cite{Tserkovnyak2002,Tserkovnyak2005RMP,Brataas2012} into a much broader class of centrosymmetric multiorbital materials that preserve both inversion and time-reversal symmetries generates spin and orbital swapping, which we trace to the locked spin and orbital textures. Orbital textures are ubiquitous in multiorbital materials owing to crystal-field effects and momentum-dependent orbital hybridization~\cite{Go2018,Lee2021,Yang2024,Sahu2021,Sahu2024,Canonico2020,Cysne2021,Costa2023,Cysne2023,Fukami2026,RyooPark2017,Hayashi2024,Go2025OrbitalPumping,Keller2025,Wang2025,Orbital_Pumping2026,Go2021EPL,JoReview2024,AtenciaReview2024,Ando2025,Cysne2025}. Through the atomic spin-orbit interaction, these orbital textures give rise to momentum-dependent spin textures nearly collinear with their orbital counterparts in centrosymmetric crystals. This enables spin pumping to generate additional swapped spin and orbital currents, as illustrated in Fig.~\ref{spin_injection}. For a saturation magnetization parallel to the interface ($\mathbf{M}_s\parallel\hat{\mathbf{y}}$), the conventional spin current $J_{S_y}^{z}$ is accompanied by a swapped current $J_{S_z}^{y}$, whereas for a saturation magnetization normal to the interface ($\mathbf{M}_s\parallel\hat{\mathbf{z}}$), the conventional spin current $J_{S_z}^{z}$ additionally generates $J_{S_x}^{x}$ and $J_{S_y}^{y}$. Beyond swapped spin currents, this effect also generates swapped orbital currents by a mechanism distinct from orbital pumping~\cite{Hayashi2024,Go2025OrbitalPumping,Keller2025,Wang2025,Orbital_Pumping2026,Pezo2025,Han2025,Ning2026} and diffusion~\cite{Ning2025}. 
We interpret these phenomena as the interconversion of spin and orbital currents,  mediated by spin-orbital textures and governed by fourth-rank polar tensors from Neumann's principle~\cite{Nye1985,Seemann2015}. Such swapping responses are universally allowed by crystalline symmetry: performing the symmetry analysis for group-IV semiconductors (Ge and Si) and beyond (e.g., GaAs, GaN, CuInSe$_2$, and SnSe) establishes the general symmetry-allowed nature of spin swapping.

\textit{General theory}.---We illustrate the general principle of angular-momentum swapping in the setup in Fig.~\ref{spin_injection} and show how the ferromagnetic resonance (FMR) in the ferromagnet pumps swapped spin and orbital currents into adjacent multiorbital materials.
We consider a multiband Hamiltonian in centrosymmetric materials, described by the matrix ${\cal H}_f({\bf k})$ under the orbit-spin product basis $|O\in\{O_1,O_2,O_3,\cdots\}\rangle\otimes |s\in\{\uparrow,\downarrow\}\rangle$. The eigenstate
$\varphi_{n,{\bf k}}$ with the dispersion $\varepsilon_n({\bf k})$ for band $n$ obeys
${\cal H}_f({\bf k})\varphi_{n,{\bf k}}=\varepsilon_n({\bf k})\varphi_{n,{\bf k}}$, viz., the unitary transformation $U_f({\bf k})=(\varphi_{1,{\bf k}},\varphi_{2,{\bf k}},\cdots)$ diagonalizes the Hamiltonian matrix according to $U^{\dagger}_f({\bf k}){\cal H}_f({\bf k})U_f({\bf k})={\rm diag}(\varepsilon_1({\bf k}),\varepsilon_2({\bf k}),\cdots)$. We shall take the Luttinger Hamiltonian~\cite{Luttinger1955,Luttinger1956} in the valence band as a prototypical example, so  $\varepsilon^h_n({\bf k})=-\varepsilon_n(-{\bf k})$ denotes the dispersion of holes in band $n$ with the wave vector ${\bf k}$, which are all doubly degenerate due to the Kramers degeneracy~\cite{kramers,Winkler2003}. 
Our theory can be routinely extended to the conduction band by particle-hole transmutation.
With $\hat{h}_n({\bf k})$ denoting the hole annihilation operator in band $n$, the free Hamiltonian of holes
\begin{align}
    \hat{H}_f=\sum_n\sum_{\bf k}\varepsilon^h_n({\bf k})\hat{h}^\dagger_n({\bf k})\hat{h}_n({\bf k}).
    \label{H_free}
\end{align}

The spin and orbital textures are defined under the eigenstates of multiorbital materials.
The matrices for the spin operator under the basis ${|O_n\rangle\otimes|s_a\rangle}$ are diagonal in orbital indices and are independent of wave vectors, which, in terms of the Pauli matrices ${\pmb \sigma}$, are expressed by
$\pmb{\Sigma}|_{\{O_n\otimes s_a\}\{O_m\otimes s_b\}}=\delta_{nm}\pmb{\sigma}_{s_as_b}$. 
In the spin operator $\hat{\bf S}=({1}/{V})\sum_{n,m}\sum_{\bf k}\pmb{\cal S}^h_{{\bf k}}|_{nm}\hat{h}^\dagger_{n,{\bf k}}\hat{h}_{m,{\bf k}}$ with $V$ being the crystal volume, $\pmb{\cal S}^h_{\bf k}=-(U_f^\dagger(-{\bf k})\pmb{\Sigma} U_f(-{\bf k}))^T/2$
denotes the spin matrix of holes.
The orbital matrix with elements $\pmb{\cal L}|_{mn}$ in the basis ${|O_n\rangle\otimes|s_a\rangle}$ is derived in the Supplemental Material (SM)~\cite{supplement}.
By explicit calculation of the orbital matrices in the Luttinger model, we show that the SOC renormalizes $\pmb{\cal L}|_{mn}$ away from the atomic orbital angular momentum (OAM) used in previous calculations~\cite{Bernevig2005,Sohn2024} (see  SM~\cite{supplement}).
In the OAM operator $\hat{\bf L}=(1/V)\sum_{n,m}\sum_{\bf k}\pmb{\cal L}^h_{{\bf k}}|_{nm}\hat{h}^\dagger_{n,{\bf k}}\hat{h}_{m,{\bf k}}$, $\pmb{\cal L}^h_{{\bf k}}=-(U_f^\dagger(-{\bf k})\pmb{\cal L} U_f(-{\bf k}))^T$ represents the OAM matrix of holes.  The spin $\pmb{\cal S}^h_{\bf k}$ and OAM $\pmb{\cal L}^h_{\bf k}$ matrices describe the spin and orbital structures, in which diagonal elements denote the spin and orbital angular momentum carried by holes of associated bands, while off-diagonal terms are correlations between different bands; their dependence on ${\bf k}$ defines the spin and orbital textures.

At the interface in the $x$-$y$ plane with $z=0$ between ferromagnet and multiorbital materials, the hole spin interacts with the magnetization via the interfacial exchange interaction.
We focus on the pumping by the FMR in the microwave-driven ferromagnet.  That is, the dynamical magnetization  ${\bf M}(t)=\sum_{\zeta=\pm}{\bf M}^{(\zeta)}e^{-i\zeta\omega t}$ precessing with frequency $\omega$ is uniform in the ferromagnet $(z<0)$, which is real by ${\bf M}^{(+)}={\bf M}^{(-)*}$. In this scenario, the system exhibits translational invariance in the $\hat{\bf x}$ and $\hat{\bf y}$ directions, so the parallel component of the wave vector $\pmb{\kappa}=k_x\hat{\bf x}+k_y\hat{\bf y}$ is conserved. The coupling between the holes and magnetization due to the interfacial exchange interaction $J_{\rm ex}$ then reads
\begin{align}
\hat{H}_{\rm int}&=J_{\rm ex}\int_V d{\bf r} \hat{\bf s}({\bf r})\cdot{\bf M}(t)\delta(z=0)\nonumber\\
&=\sum_{\zeta=\pm}\sum_{n,m}\sum_{{\pmb \kappa},k_z,k_z'}G_{{k}_z'{k}_z}^{(\zeta)}(\pmb{\kappa})|_{nm}e^{-i\zeta\omega t}\hat{h}^\dagger_{n,k_z'}(\pmb{\kappa})\hat{h}_{m,k_z}(\pmb{\kappa}).
\nonumber
\end{align}
The dynamical magnetization scatters the holes between wave-vector components $k_z$ and $k_z'$ and across different bands $m$ and $n$, governed by the amplitude 
\[
G_{k_z'k_z}^{(\zeta)}(\pmb{\kappa})=-\frac{\hbar J_{\rm ex}}{2L_z}\left(U_f^\dagger(-\pmb{\kappa},-k_z)\pmb{\Sigma}U_f(-\pmb{\kappa},-k_z')\right)^T\cdot{\bf M}^{(\zeta)},
\]
in which $L_z$ is the thickness of multiorbital materials along interface normal $\hat{\bf z}$-direction. When $k_z=k_z'$, the magnetization couples directly to the spin textures, but generally it involves more dynamics with $k_z\ne k_z'$.

\begin{figure}[htp!]
\centering
\includegraphics[width=0.43\textwidth,trim=0cm 0cm 0cm 0cm]{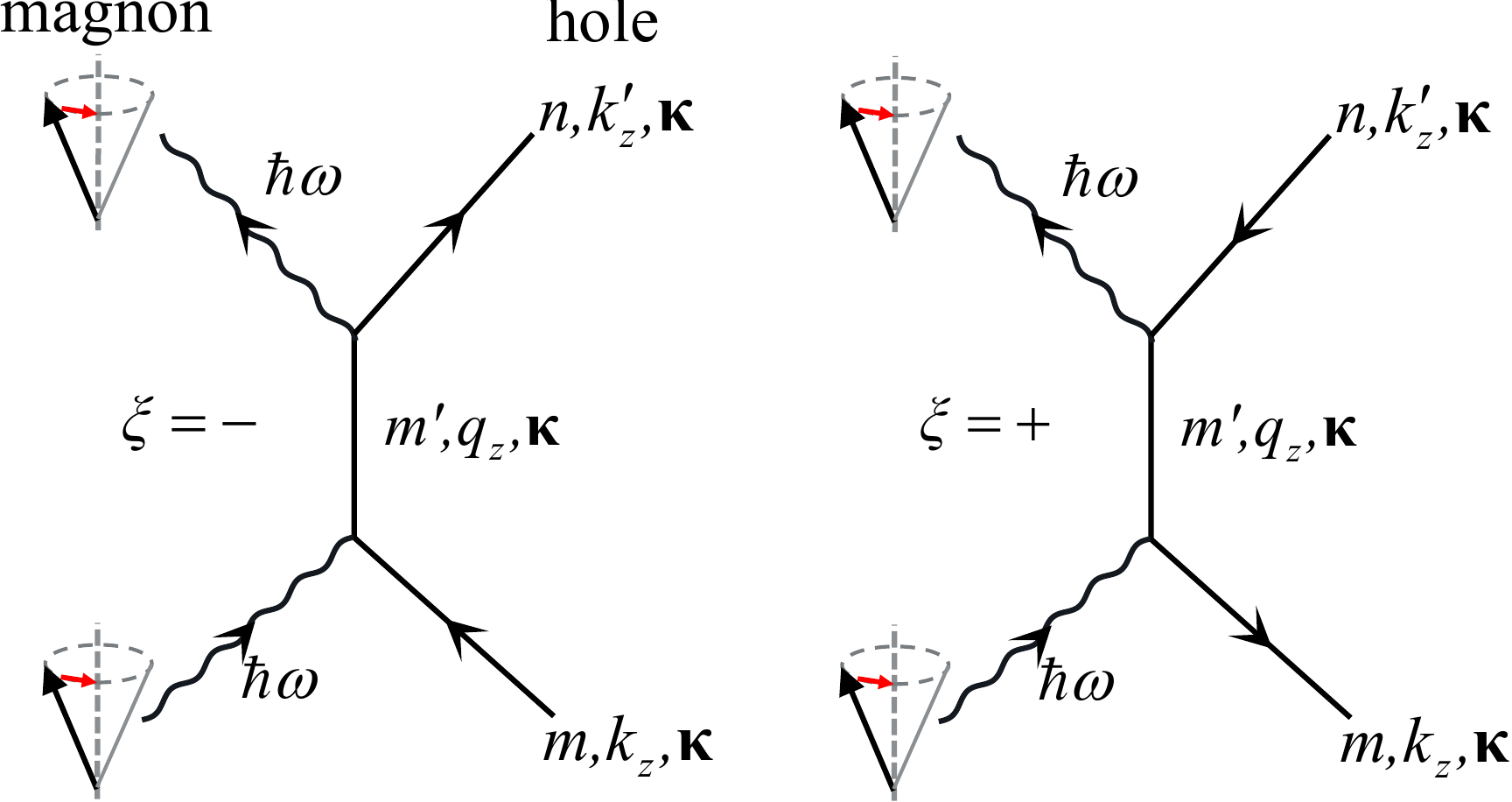}
\caption{Two typical hole-magnon scattering processes induced by the interfacial $s$-$d$ exchange interaction. The coherent magnons induce the hole coherence between different wave vectors and bands.}
\label{scattering_process}
\end{figure}

During typical scattering processes illustrated in Fig.~\ref{scattering_process}, the hole correlation $\langle \hat{h}^\dagger_{n,k_z'}(\pmb{\kappa})\hat{h}_{m,k_z}(\pmb{\kappa})\rangle$ between wave vectors $\{k_z,k_z'\}$ and bands $\{m,n\}$ is induced: the hole with wave vector $k_z$ in band $m$ first absorbs ($\zeta=-$) or emits ($\zeta=+$) a magnon of frequency $\omega$ and transfers to the intermediate state with wave vector $q_z$ in the band $m'$ with energy $\varepsilon^h_{m'}(\pmb{\kappa},q_z)=\varepsilon^h_{m}(\pmb{\kappa},k_z)-\zeta\hbar\omega$, which then emits or absorbs a magnon and arrives at the final state with wave vector $k'_z$ in band $n$, governed by the energy conservation $\varepsilon^h_{m'}(\pmb{\kappa},q_z)=\varepsilon^h_{n}(\pmb{\kappa},k'_z)-\zeta\hbar\omega$. 
For the pumped waves propagating to the positive half space $z>0$, we need to concern the wave vectors $\{k_z,k_z'\}$ with positive group velocities, viz., $k_z=q^{(\zeta)}_{z,mm'}$ and $k_z'=q_{z,nm'}^{(\zeta)}$ are roots of $\varepsilon^h_{m}(\pmb{\kappa},q^{(\zeta)}_{z,mm'})=\varepsilon^h_{n}(\pmb{\kappa},q^{(\zeta)}_{z,nm'})=\varepsilon^h_{m'}(\pmb{\kappa},q_z)+\zeta\hbar\omega$ with positive group velocities  $\partial_{k_z}\varepsilon^h_{m,k_z}(\pmb{\kappa})|_{k_z=q_{z,mm'}^{(\zeta)}}>0$ and $\partial_{k_z}\varepsilon^h_{n,k_z}(\pmb{\kappa})|_{k_z=q_{z,nm'}^{(\zeta)}}>0$.
The quantum coherence between the hole initial state $|m,q_{z,mm'}^{(\zeta)}\rangle$ and final state $|n,q_{z,nm'}^{(\zeta)}\rangle$ induced by magnon-assisted scattering channels gives rise to the pumped spin and orbital currents. Particularly, for spatially uniform currents, only contributions with $q_{z,mm'}^{(\zeta)}=q_{z,nm'}^{(\zeta)}$ survive, which, together with the energy conservation, requires the bands $m$ and $n$ in the established coherence to be degenerate. 

We substantiate the expectation by finding the driven density matrix of holes (see SM~\cite{supplement}), which accounts for the spin and orbital dynamics driven by the dynamical exchange magnetic field $J_{\rm ex}{\bf M}(t)$. In the ballistic regime, we obtain the DC spin ``$S$" and orbital ``$L$" currents at $z>0$ pumped by the FMR 
\begin{align}
{J}^{i}_{S_\alpha/L_\alpha}
    &=\frac{\hbar^2J_{\rm ex}^2}{4(2\pi)^3}\sum_{\zeta=\pm}\sum_{ \{m, n\}}\sum_{ m'}\int d{\pmb \kappa} dq_z{\mathcal{J}_{S/L,\alpha}^{h,i}\left(\tilde{q}_z,{\pmb \kappa}\right)|_{ n m}}\nonumber\\
    &\times \bigg(\varphi^*_{ m',-q_z}(-{\pmb \kappa})\pmb{\Sigma}\cdot{\bf M}^{(\zeta)}\varphi_{ m,-\tilde{q}_z}(-{\pmb \kappa})\bigg)\nonumber\\
    &\times\left(\varphi^*_{ n,-\tilde{q}_z}(-{\pmb \kappa})\pmb{\Sigma}\cdot{\bf M}^{(\zeta)*}\varphi_{ m',-q_z}(-{\pmb \kappa})\right)\nonumber\\
&\times\frac{f\left(\varepsilon^h_{ m',q_z}({\pmb \kappa})\right)-f\left(\varepsilon^h_{ m',q_z}({\pmb \kappa})+\zeta\hbar\omega\right)}{\left(\partial_{\tilde{q}_z}\varepsilon^h_{ m,\tilde{q}_z}(\pmb{\kappa})\right)^2}.
    \label{pumped_current}
\end{align}
${\cal J}^{h,i}_{{\cal O},\alpha}({\bf k})=-[U^\dagger_f(-{\bf k})\left\{{\cal O}_\alpha,{ \cal V}_i(-{\bf k})\}U_f(-{\bf k})\right]/2$ denotes the spin- and orbital-current textures for ${\cal O}_\alpha=\{\Sigma_\alpha/2,{\cal L}_\alpha\}$, which are crucial for generating spin and orbital swapping. The summation over intermediate hole states $\{q_z,m'\}$ in Fig.~\ref{scattering_process} is implied; the braces in $\{m,n\}$ indicate they are degenerate-band pairs, the wave vector $\tilde{q}_z\equiv q_{z,mm'}^{(\zeta)}=q_{z,nm'}^{(\zeta)}$ indicates the energy conservation in the scattering process and positive group velocities in the transport, and $f(\varepsilon^h_{ m,{\bf k}})=1/(e^{(\varepsilon^h
_{ m,\mathbf{k}}-\mu)/(k_{B}T)}+1)$ is the equilibrium Fermi-Dirac distribution of holes at temperature $T$ and chemical potential $\mu$.

 In multiorbital materials without symmetry-protected band degeneracy, such as Rashba systems with broken inversion $\cal P$ symmetry~\cite{Sadjina2012,Borge2017,Li2019,Chen2015SpinPumpingSOC} and ferromagnetic metals with broken time-reversal $\cal T$ symmetry~\cite{Park2022}, electron bands are nondegenerate at generic momenta, forcing $m=n$ in Eq.~\eqref{pumped_current}. By contrast, in centrosymmetric systems preserving time-reversal symmetry, the combined $\cal{PT}$ symmetry enforces Kramers degeneracy, giving rise to twofold-degenerate subspaces at every wave vector~\cite{kramers,Zhang2014}. 
 We then choose \textit{a proper band basis} within the degenerate subspace to render the spin-current texture matrix  ${\mathcal{J}_{S,\alpha}^{h,i}({\bf k})|_{ n m}}=\delta_{mn}({\cal S}^{h}_{{\bf k},\alpha}{\cal V}_i (-{\bf k}))|_{nn}$ diagonal, which also requires $m=n$.  Thus, the pumping process can be understood as a magnon-induced, non-equilibrium redistribution $\delta n_{n}({\bf k})$ of holes in band $n$ across different wave vectors, giving rise to a tensorial spin current due to the underlying spin textures (see below). The same analysis applies to the orbital current, replacing the spin texture with the orbital one.

\textit{Group-IV semiconductors}.---Group-IV semiconductors provide a prototypical model system with orbital textures originating from multiple $p$-orbitals in valence bands~\cite{Luttinger1955,Luttinger1956,Kane1957,Lawaetz1971,Winkler2003,Culcer2004,Kane_2}. Orbital transport in these materials has been of particular interest because an intrinsic orbital Hall effect was originally predicted for $p$-doped Si~\cite{Bernevig2005} and has recently been investigated experimentally~\cite{Matsumoto2025}. We focus on the valence bands with three $p$-orbitals, incorporating the SOC between the $p$-orbitals and spins. When such atomic SOC is explicitly retained, the angular momentum injected by FMR is not confined to the spin sector, but interplays with the orbital sector. 
The Luttinger Hamiltonian~\cite{Luttinger1955,Luttinger1956} around the $\Gamma$ point under the basis of $|P_x,\uparrow\rangle$, $|P_x,\downarrow\rangle$, $|P_y,\uparrow\rangle$, $|P_y,\downarrow\rangle$, $|P_z,\uparrow\rangle$, and $|P_z,\downarrow\rangle$ reads 
\begin{align}
\hat{H}_f({\bf r})&=
A\hat{k}^2-(A-B)\sum_i\hat{k}_i^2{\cal I}_i^2-\frac{C}{2}\sum_{i\neq j}\hat{k}_i\hat{k}_j\{{\cal I}_i,{\cal I}_j\}\nonumber\\
&+({\lambda}/{3})\left(\boldsymbol{\cal I}\cdot\boldsymbol{\Sigma}-E_{6\times6}\right),
\label{free_Hamiltonian_r}
\end{align}
in which  the coefficients $\{A,B,C\}$~\cite{Luttinger1955,Luttinger1956}, the atomic orbital
$\pmb{\cal I}$, and the spin-${\pmb \Sigma}$ matrices are detailed in SM~\cite{supplement}.
 $E_{6\times 6}$ is the identity matrix, and $\lambda$ denotes the energy splitting at the $\Gamma$ point between the spin-split-off band and heavy-hole/light-hole bands.

In Ge, the strong SOC with $\lambda=290$~meV~\cite{Lawaetz1971} separates the spin-split-off band from the heavy-hole and light-hole bands, while the heavy-hole and light-hole dispersions split away from the $\Gamma$ point due to the crystal field effect. All these bands remain twofold degenerate at generic momenta, enforced by the combined $\mathcal{PT}$ symmetry. With a light doping, the spin-split-off band is negligibly populated and thus contributes little to the pumped current. Furthermore, the current contribution in Eq.~\eqref{pumped_current} is inversely proportional to the square of the group velocity $\left[\partial_{\tilde{q}_z}\varepsilon^h_{m,\tilde{q}_z}(\pmb{\kappa})\right]^2$. Because the heavy-hole group velocity is much smaller than that of the light holes, the heavy-hole states dominate the pumping response.
While our calculation includes all contributions,  we here trace the spin- and orbital-swapped currents to be induced by spin and orbital textures of heavy-hole states.

\begin{figure}[htp!]
\centering
\includegraphics[width=0.42\textwidth,trim=0cm 0cm 0cm 0cm]{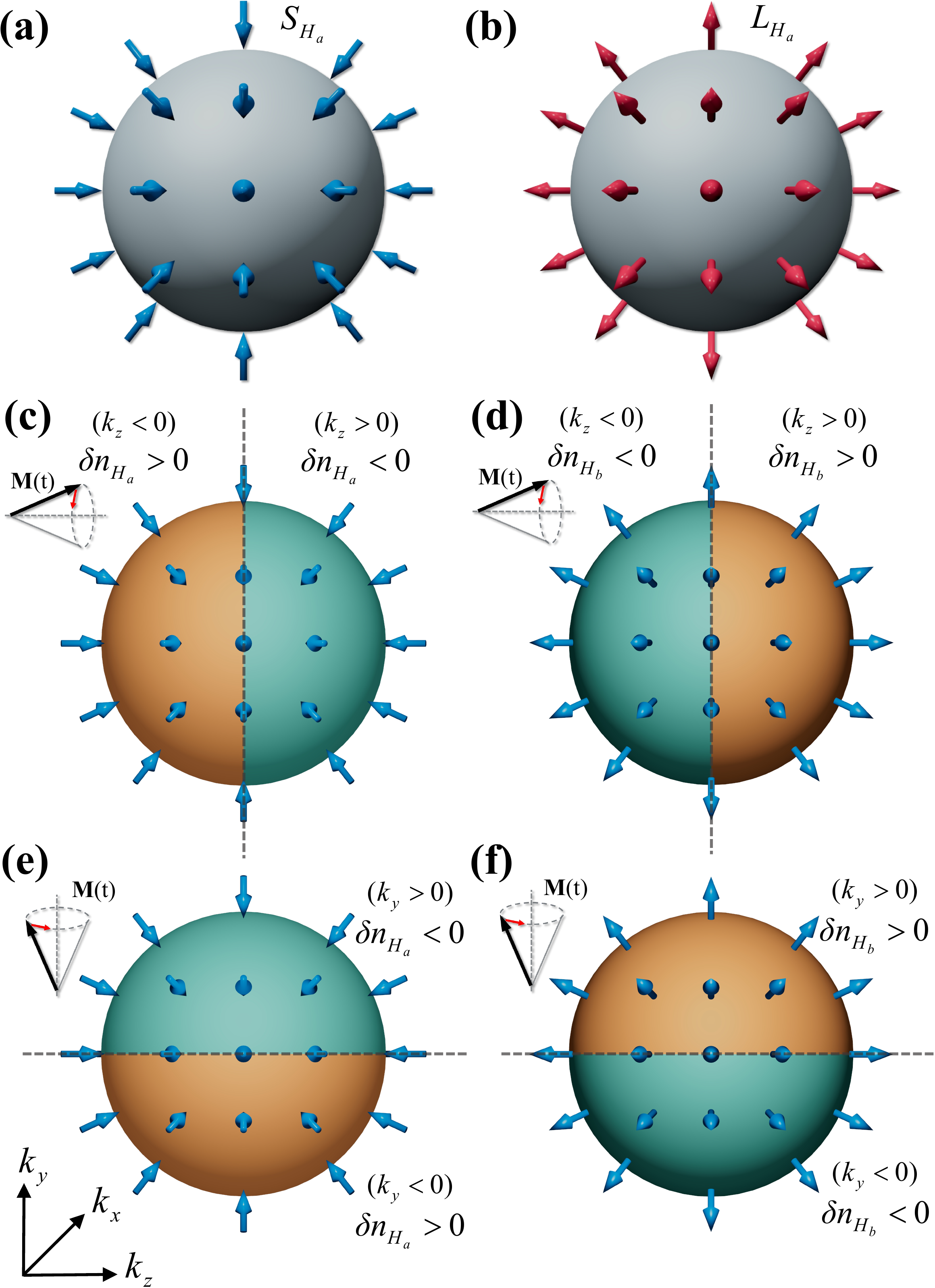}
\caption{Spin and orbital textures \textit{vs}. scattering-induced population changes of holes in the degenerate heavy-hole manifold.
(a) and (b) plot the spin $\pmb{\cal S}^h_{{\bf k}}$ and orbital $\pmb{\cal L}^h_{{\bf k}}$ textures of holes in one heavy-hole band $H_a$. The spin texture is nearly radially inward, whereas the OAM texture is nearly radially outward. (c) and (d) illustrate the changes in hole population $\delta n_{H_a}$ and $\delta n_{H_b}$ in two hole bands when the magnon spin is normal to the interface. (e) and (f) show the other situation when the magnon spin is parallel to the interface. Orange and green regions indicate an increase ($\delta n>0$) and a decrease ($\delta n<0$) of hole population.}
\label{texture}
\end{figure}

By the strong SOC $\lambda$, the spin and orbital textures of heavy-hole bands $\{H_a,H_b\}$ are locked to the wave-vector direction. Figure~\ref{texture}(a) and (b) illustrate the spin and orbital textures in one heavy-hole band $H_a$, showing that the spin texture is nearly radially inward, whereas the orbital texture is nearly radially outward. Both textures are reversed for the other heavy-hole band $H_b$. 
Magnon-assisted scattering preferentially increases the hole populations with spins aligned with the saturation magnetization, leading to a momentum-dependent redistribution and imbalance of holes.

When the saturation magnetization is normal to the interface [Fig.~\ref{texture}(c) and (d)], the two heavy-hole bands develop opposite population changes: $\delta n_{H_a}(k_z<0)>0$ and $\delta n_{H_a}(k_z>0)<0$ in $H_a$ [Fig.~\ref{texture}(c)], with opposite changes in $H_b$ [Fig.~\ref{texture}(d)]. Since only states with positive group velocities ($k_z>0$ for the nearly parabolic heavy-hole bands) contribute to the currents pumped into the $z>0$ region, the depletion $\delta n_{H_a}(k_z>0)<0$ in $H_a$ and the enhancement $\delta n_{H_b}(k_z>0)>0$ in $H_b$, combined with their opposite spin textures [Fig.~\ref{texture}(a)], constructively generate the conventional spin current $J^z_{S_z}$ together with swapped components $J^x_{S_x}$ and $J^y_{S_y}$.
For an in-plane saturation magnetization ($\mathbf{M}_s\parallel\hat{\mathbf{y}}$) [Fig.~\ref{texture}(e) and (f)], the population changes instead occur across $k_y=0$: $\delta n_{H_a}(k_y>0)<0$ and $\delta n_{H_a}(k_y<0)>0$ in $H_a$ [Fig.~\ref{texture}(e)], with opposite changes in $H_b$ [Fig.~\ref{texture}(f)]. 
Within the $k_z>0$ hemisphere, the intraband spin-current textures convert the $k_y$ imbalance within each band into both the conventional current $J^z_{S_y}$ and the swapped current $J^y_{S_z}$, with $H_b$ contributing identically because both its population changes and spin texture are reversed [Fig.~\ref{texture}(e) and (f)]. The same mechanism generates orbital currents through the corresponding orbital-current textures [Fig.~\ref{texture}(b)]. Thus, the interplay between spin and orbital textures and magnon-induced hole scattering (Fig.~\ref{scattering_process}) determines the allowed spin and orbital current components.

The spin and orbital currents injected into bulk Ge by the FMR of yttrium iron garnet (YIG)~\cite{Kajiwara2010,RojasSanchez2013,Santos2024GeOHE} are calculated numerically. The Luttinger-Hamiltonian parameters for Ge are  $A=-1.85\times10^{-37}~\mathrm{J}\cdot\mathrm{m}^2$, $B=-2.96\times10^{-38}~\mathrm{J}\cdot\mathrm{m}^2$, and $C=-2.08\times10^{-37}~\mathrm{J}\cdot\mathrm{m}^2$ \cite{Luttinger1956,Lawaetz1971}. 
The calculated parameters for the orbital matrix are given in End Matter. The applied magnetic field $\mu_0H_0=50~\mathrm{mT}$ corresponds to an FMR frequency $\omega =2\pi\times1.45$~GHz. The driven transverse magnetization is $0.2M_s$, with $\mu_0 M_s = 0.177~\mathrm{T}$ for YIG. The exchange coupling $J_{\rm ex} = 5.4\times10^{-3}~\mathrm{A}^{-1}\cdot\mathrm{s}^{-1}$ corresponds to an exchange spin splitting $\sim 1~\mathrm{meV}$.

\begin{figure}[htp!]
\centering
\includegraphics[width=0.235\textwidth,trim=0cm 0cm 0cm 0cm]{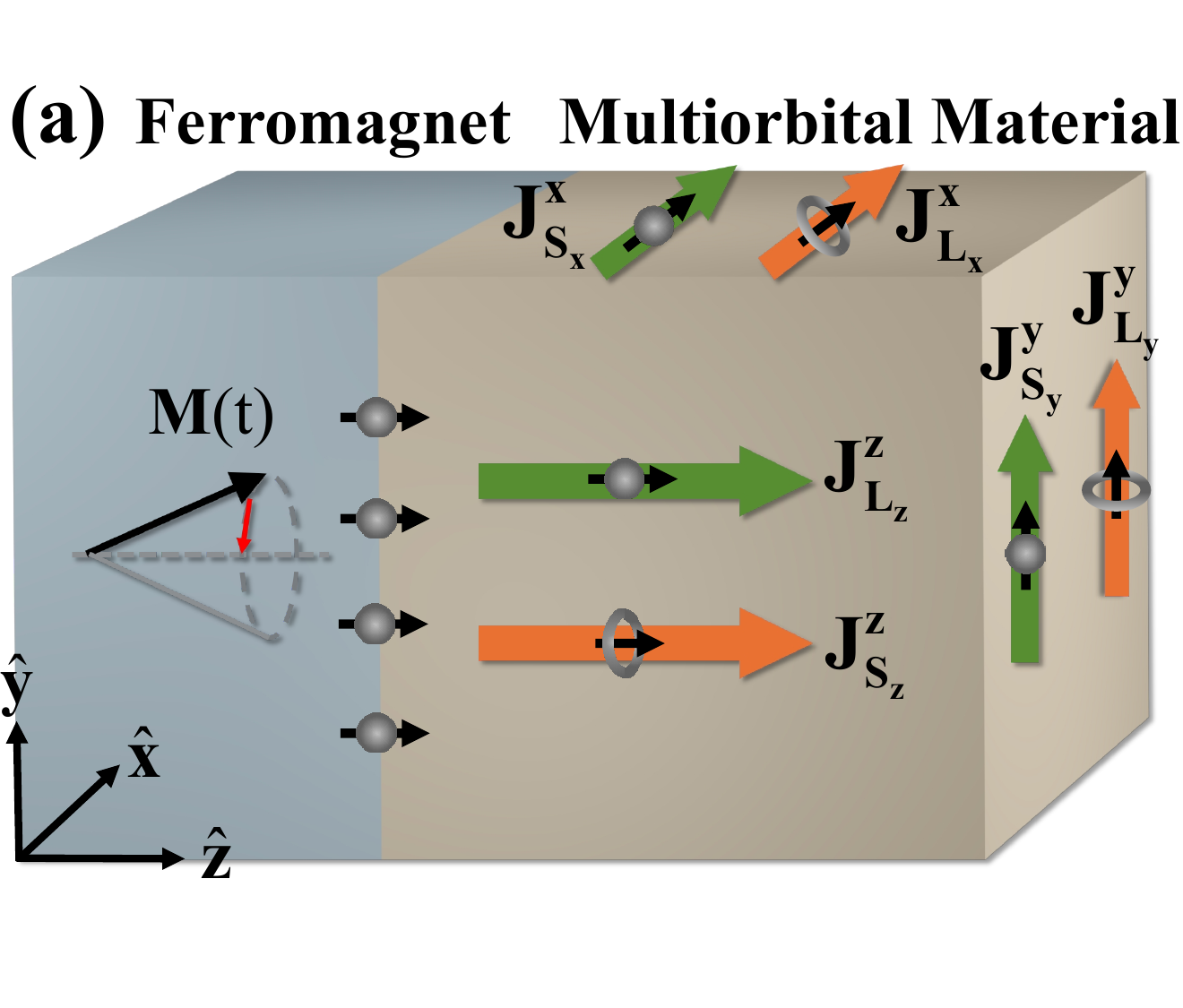}
\includegraphics[width=0.235\textwidth,trim=0cm 0cm 0cm
0cm]{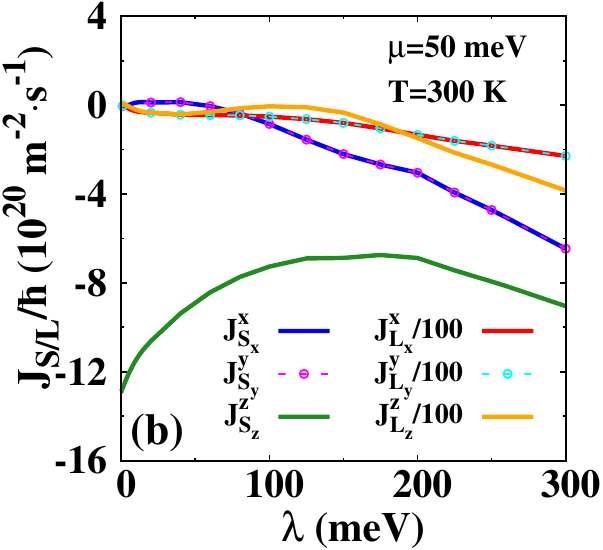}
\includegraphics[width=0.235\textwidth,trim=0cm 0cm 0cm 0cm]{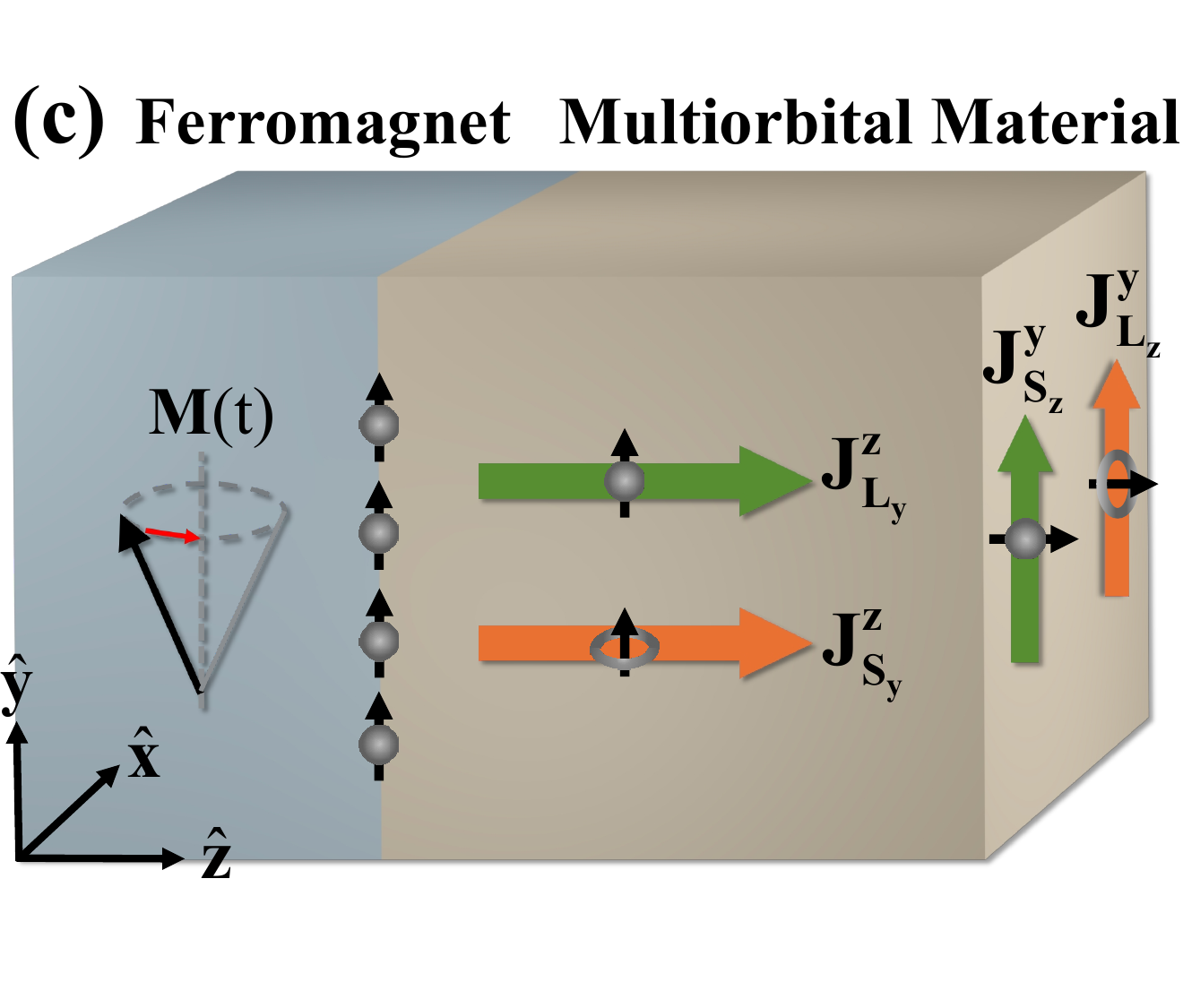}
\includegraphics[width=0.235\textwidth,trim=0cm 0cm 0cm 0cm]{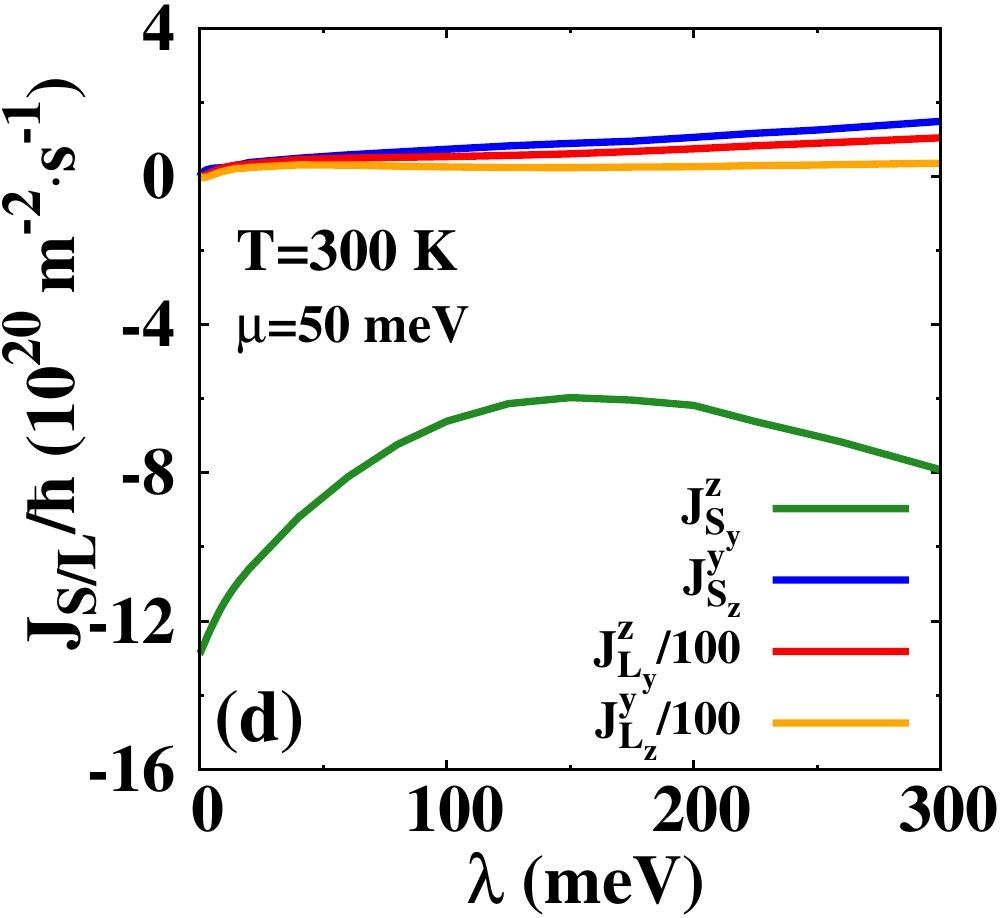}
\caption{Pumped spin and orbital currents in Ge by treating the SOC strength $\lambda$ as a parameter for different magnetization drivings. (a) and (b) show the first scenario of spin and orbital swapping when the saturation magnetization is normal to the interface. (c) and (d) show the second scenario when the saturation magnetization is parallel to the interface. }
\label{swapping_current}
\end{figure}

Figure~\ref{swapping_current} plots all non-zero spin-current and orbital-current components as a function of SOC strength $\lambda$ (noting $\lambda=290$~meV for Ge~\cite{Lawaetz1971}) at room temperature $T=300~\mathrm{K}$ and chemical potential $\mu=50~\mathrm{meV}$ in lightly doped regime.  
When the saturation magnetization is normal to the interface of YIG$|$Ge, increasing the SOC strength gradually induces the swapped spin currents $J^x_{S_{x}}$ and $J^y_{S_{y}}$, as well as the orbital-swapping currents $J^x_{L_{x}}$, $J^y_{L_{y}}$, and $J^z_{L_{z}}$, in addition to the primary longitudinal spin current $J^z_{S_z}$ [Fig.~\ref{swapping_current}(a) and (b)]. The spin swapping in this configuration achieves a high efficiency $\gtrsim 50\%$. Notably, $J^x_{S_x}$ and $J^y_{S_y}$ have equal magnitudes, as do $J^x_{L_x}$ and $J^y_{L_y}$. 
These relations are dictated by the cubic crystal symmetry of the $O_h$ point group (see End Matter). When the saturation magnetization (along $\hat{\bf y}$) is parallel to the interface, increasing the SOC strength induces the swapped spin current $J^y_{S_{z}}$ and orbital currents $J^y_{L_{z}}$ and $J^z_{L_{y}}$, in addition to the primary spin current $J^z_{S_y}$ [Fig.~\ref{swapping_current}(c) and (d)].  The spin swapping efficiency is $\gtrsim 20\%$ in this case. 

The pumped orbital current in Ge is substantial, whereas in Si it is much weaker (see SM~\cite{supplement}). This enhancement is fully consistent with previous studies~\cite{RangKelly2024,ValetPRB2025,ValetPRL2025}. Indeed, in degenerate bands protected by ${\cal PT}$-symmetry, the surviving orbital current is an intraband transport process~\cite{ValetPRL2025}, so its magnitude is set by the orbital magnetic moment carried by each carrier: a hole in Ge carries about $13\hbar$ of OAM (Table~\ref{tab3}, End Matter), vastly larger than the spin value of $\hbar/2$. For comparison, if the orbital and spin magnetic moments are taken to be equal, the orbital current falls below the spin current.

Spin pumping into nonmagnetic centrosymmetric multiorbital materials generates swapped spin and orbital currents through an explicit spin$\,\to\,$orbital$\,\to\,$spin intermediate channel, governed solely by crystalline symmetry rather than by magnetic order~\cite{Park2022} or spin-band splitting. Such spin swapping exists in a wide range of materials beyond group-IV semiconductors (see End Matter).

\textit{Discussion and conclusion}.---In summary, we uncover an orbital-mediated mechanism of spin and orbital swapping driven by spin pumping to centrosymmetric multiorbital materials. Spin-orbit-entangled textures redirect the injected angular momentum into swapped spin and orbital currents, without requiring spin-band splitting or extrinsic spin-orbit scattering. 
Microscopic calculations for ferromagnet$|$group-IV semiconductors demonstrate the high efficiency of this mechanism, $ 20\%$ to $ 50\%$, which depends on SOC and magnetization orientation.  A fourth-rank spin-orbital conversion tensor establishes the crystalline-symmetry constraints on these responses beyond the specific Hamiltonian model. Fabrication of heavy-metal electrodes on the lateral faces of the crystal favors future detection of the predicted effect. By identifying orbital textures as a source of angular-momentum redirection, our work provides a framework for controlling coupled spin and orbital transport through band structure and crystal symmetry and opens a new way for the realization of angular-momentum swapping with unexplored material choices.

\section{acknowledgments}
This work was supported by the National Natural Science Foundation of China under Grants No.~12374109 and 12474100, as well as the National Key Research and Development Program of China under Grant No.~2023YFA1406600. H.W. acknowledges the support of Swiss National Science Foundation (Grant No.~10004526). We thank Gerrit Bauer for useful discussions.

\bibliography{refs}

@article{Lifshits2009,
  author  = {Lifshits, Maria B. and Dyakonov, Michel I.},
  title   = {Swapping Spin Currents: Interchanging Spin and Flow Directions},
  journal = {Physical Review Letters},
  volume  = {103},
  number  = {18},
  pages   = {186601},
  year    = {2009},
  doi     = {10.1103/PhysRevLett.103.186601}
}

@article{Sadjina2012,
  author  = {Sadjina, Severin and Brataas, Arne and Mal'shukov, A. G.},
  title   = {Intrinsic Spin Swapping},
  journal = {Physical Review B},
  volume  = {85},
  number  = {11},
  pages   = {115306},
  year    = {2012},
  doi     = {10.1103/PhysRevB.85.115306}
}

@article{Shen2015,
  author  = {Shen, Ka and Raimondi, R. and Vignale, G.},
  title   = {Spin current swapping and Hanle {spin Hall effect} in a two-dimensional electron gas},
  journal = {Physical Review B},
  volume  = {92},
  number  = {3},
  pages   = {035301},
  year    = {2015},
  doi     = {10.1103/PhysRevB.92.035301}
}

@article{Saidaoui2015,
  author  = {Saidaoui, Hamed Ben Mohamed and Otani, YoshiChika and Manchon, Aurelien},
  title   = {Crossover between spin swapping and Hall effect in disordered systems},
  journal = {Physical Review B},
  volume  = {92},
  number  = {2},
  pages   = {024417},
  year    = {2015},
  doi     = {10.1103/PhysRevB.92.024417}
}

@article{Saidaoui2016,
  author  = {Saidaoui, Hamed Ben Mohamed and Manchon, Aurelien},
  title   = {Spin-Swapping Transport and Torques in Ultrathin Magnetic Bilayers},
  journal = {Physical Review Letters},
  volume  = {117},
  number  = {3},
  pages   = {036601},
  year    = {2016},
  doi     = {10.1103/PhysRevLett.117.036601}
}

@article{Espedal2017,
  author  = {Espedal, Camilla and Lange, Peter and Sadjina, Severin and Mal'shukov, A. G. and Brataas, Arne},
  title   = {{Spin Hall effect} and spin swapping in diffusive superconductors},
  journal = {Physical Review B},
  volume  = {95},
  number  = {5},
  pages   = {054509},
  year    = {2017},
  doi     = {10.1103/PhysRevB.95.054509}
}

@article{Borge2017,
  author  = {Borge, Juan and Tokatly, Ilya V.},
  title   = {Ballistic spin transport in the presence of interfaces with strong spin-orbit coupling},
  journal = {Physical Review B},
  volume  = {96},
  number  = {11},
  pages   = {115445},
  year    = {2017},
  doi     = {10.1103/PhysRevB.96.115445}
}

@article{Ortiz2018,
  author  = {Ortiz Pauyac, Christian and Chshiev, Mairbek and Manchon, Aurelien and Nikolaev, Sergey A.},
  title   = {Spin Hall and Spin Swapping Torques in Diffusive Ferromagnets},
  journal = {Physical Review Letters},
  volume  = {120},
  number  = {17},
  pages   = {176802},
  year    = {2018},
  doi     = {10.1103/PhysRevLett.120.176802}
}

@article{Li2019,
  author  = {Li, Song and Shen, Ka and Xia, Ke},
  title   = {Interfacial {spin Hall effect} and spin swapping in {Fe-Au bilayers} from first principles},
  journal = {Physical Review B},
  volume  = {99},
  number  = {13},
  pages   = {134427},
  year    = {2019},
  doi     = {10.1103/PhysRevB.99.134427}
}

@article{Park2022,
  author  = {Park, Hyeon-Jong and Ko, Hye-Won and Go, Gyungchoon and Oh, Jung Hyun and Kim, Kyoung-Whan and Lee, Kyung-Jin},
  title   = {Spin Swapping Effect of Band Structure Origin in Centrosymmetric Ferromagnets},
  journal = {Physical Review Letters},
  volume  = {129},
  number  = {3},
  pages   = {037202},
  year    = {2022},
  doi     = {10.1103/PhysRevLett.129.037202}
}

@article{Cui2023,
  author  = {Cui, Shuting and Yan, Peng and Shen, Maokang and Luo, Wei and Yang, Xiaofei and Zhang, Yue},
  title   = {Spin swapping for an exchange magnon spin current},
  journal = {Physical Review B},
  volume  = {108},
  number  = {18},
  pages   = {L180402},
  year    = {2023},
  doi     = {10.1103/PhysRevB.108.L180402}
}

@article{Kamra2024,
  author  = {Kamra, Lina Johnsen and Linder, Jacob},
  title   = {Inverse Spin-Hall Effect and Spin Swapping in Spin-Split Superconductors},
  journal = {Physical Review Letters},
  volume  = {132},
  number  = {22},
  pages   = {226002},
  year    = {2024},
  doi     = {10.1103/PhysRevLett.132.226002}
}

@article{Ning2025,
  author  = {Ning, Xiaobai and Pezo, A. and Kim, Kyoung-Whan and Zhao, Weisheng and Lee, Kyung-Jin and Manchon, Aurelien},
  title   = {Orbital Diffusion, Polarization, and Swapping in Centrosymmetric Metals},
  journal = {Physical Review Letters},
  volume  = {134},
  number  = {2},
  pages   = {026303},
  year    = {2025},
  doi     = {10.1103/PhysRevLett.134.026303}
}

@article{Kim2020,
  author  = {Kim, Heeman and Karube, Shutaro and Borge, Juan and Kim, Junyeon and Kondou, Kouta and Otani, YoshiChika},
  title   = {Evidence for spin swapping from modulation of transverse resistance in magnetic heterostructures with {Rashba interface}},
  journal = {Applied Physics Letters},
  volume  = {116},
  number  = {12},
  pages   = {122403},
  year    = {2020},
  doi     = {10.1063/1.5143955}
}

@article{Fu2022,
  author  = {Fu, Qingwei and Liang, Like and Wang, Wenqiang and Yang, Liupeng and Zhou, Kaiyuan and Li, Zishuang and Yan, Chunjie and Li, Liyuan and Li, Haotian and Liu, Ronghua},
  title   = {Observation of nontrivial spin-orbit torque in single-layer ferromagnetic metals},
  journal = {Physical Review B},
  volume  = {105},
  number  = {22},
  pages   = {224417},
  year    = {2022},
  doi     = {10.1103/PhysRevB.105.224417}
}

@article{Hazra2023,
  author  = {Hazra, Binoy K. and Pal, Banabir and Jeon, Jae-Chun and
             Neumann, Robin R. and G{\"o}bel, B{\"o}rge and
             Grover, Bharat and Deniz, Hakan and Styervoyedov, Andriy and
             Meyerheim, Holger and Mertig, Ingrid and
             Yang, See-Hun and Parkin, Stuart S. P.},
  title   = {Generation of out-of-plane polarized spin current by spin swapping},
  journal = {Nature Communications},
  volume  = {14},
  pages   = {4549},
  year    = {2023},
  doi     = {10.1038/s41467-023-39884-6}
}

@article{Kim2025,
  author  = {Kim, Kitae and Shim, Woo Young and Choe, Sug-Bong},
  title   = {Huge self-spin swapping effect with asymmetric spin-sink structures},
  journal = {NPG Asia Materials},
  volume  = {17},
  pages   = {16},
  year    = {2025},
  doi     = {10.1038/s41427-025-00597-5}
}

@article{Wu2026,
  author  = {Wu, Mingxing and Kondou, Kouta and Chen, Taishi and Nakatsuji, Satoru and Otani, YoshiChika},
  title   = {Identifying the origin of out-of-plane spin polarization in the noncollinear antiferromagnet {Mn3Ge}},
  journal = {Physical Review B},
  volume  = {114},
  number  = {2},
  pages   = {L020403},
  year    = {2026},
  doi     = {10.1103/kpj1-yrwh}
}

@article{Ryu2022,
  author  = {Ryu, Jeongchun and Thompson, Ryan and Park, Jae Yeol
             and Kim, Seok-Jong and Choi, Gaeun and Kang, Jaimin
             and Jeong, Han Beom and Kohda, Makoto and Yuk, Jong Min
             and Nitta, Junsaku and Lee, Kyung-Jin and Park, Byong-Guk},
  title   = {Efficient spin--orbit torque in magnetic trilayers
             using all three polarizations of a spin current},
  journal = {Nature Electronics},
  volume  = {5},
  number  = {4},
  pages   = {217--223},
  year    = {2022},
  doi     = {10.1038/s41928-022-00735-9}
}

@article{Wang2024,
  author  = {Wang, Fei and Shi, Guoyi and Kim, Kyoung-Whan
             and Park, Hyeon-Jong and Jang, Jae Gwang
             and Tan, Hui Ru and Lin, Ming and Liu, Yakun
             and Kim, Taeheon and Yang, Dongsheng
             and Zhao, Shishun and Lee, Kyusup
             and Yang, Shuhan and Soumyanarayanan, Anjan
             and Lee, Kyung-Jin and Yang, Hyunsoo},
  title   = {Field-free switching of perpendicular magnetization
             by two-dimensional {PtTe$_2$}/{WTe$_2$}
             van der Waals heterostructures with high
             spin Hall conductivity},
  journal = {Nature Materials},
  volume  = {23},
  number  = {6},
  pages   = {768--774},
  year    = {2024},
  doi     = {10.1038/s41563-023-01774-z}
}

@article{Tserkovnyak2002PRL,
  author  = {Tserkovnyak, Yaroslav and Brataas, Arne
             and Bauer, Gerrit E. W.},
  title   = {Enhanced {Gilbert Damping} in Thin
             Ferromagnetic Films},
  journal = {Physical Review Letters},
  volume  = {88},
  number  = {11},
  pages   = {117601},
  year    = {2002},
  doi     = {10.1103/PhysRevLett.88.117601}
}

@article{Tserkovnyak2005RMP,
  author  = {Tserkovnyak, Yaroslav and Brataas, Arne
             and Bauer, Gerrit E. W. and Halperin, Bertrand I.},
  title   = {Nonlocal magnetization dynamics in
             ferromagnetic heterostructures},
  journal = {Reviews of Modern Physics},
  volume  = {77},
  number  = {4},
  pages   = {1375--1421},
  year    = {2005},
  doi     = {10.1103/RevModPhys.77.1375}
}

@article{Hirsch1999,
  author = {Hirsch, J. E.},
  title = {{Spin Hall Effect}},
  journal = {Phys. Rev. Lett.},
  volume = {83},
  pages = {1834--1837},
  year = {1999},
  doi = {10.1103/PhysRevLett.83.1834}
}

@article{Tserkovnyak2002,
  author = {Tserkovnyak, Yaroslav and Brataas, Arne and Bauer, Gerrit E. W.},
  title = {Spin pumping and magnetization dynamics in metallic multilayers},
  journal = {Phys. Rev. B},
  volume = {66},
  pages = {224403},
  year = {2002},
  doi = {10.1103/PhysRevB.66.224403}
}

@article{Sinova2015,
  author = {Sinova, Jairo and Valenzuela, Sergio O. and Wunderlich, J. and Back, C. H. and Jungwirth, T.},
  title = {{Spin Hall effects}},
  journal = {Rev. Mod. Phys.},
  volume = {87},
  pages = {1213--1260},
  year = {2015},
  doi = {10.1103/RevModPhys.87.1213}
}

@article{Manchon2019,
  author = {Manchon, A. and {\v{Z}}elezn{\'y}, J. and Miron, I. M. and Jungwirth, T. and Sinova, J. and Thiaville, A. and Garello, K. and Gambardella, P.},
  title = {Current-induced spin-orbit torques in ferromagnetic and antiferromagnetic systems},
  journal = {Rev. Mod. Phys.},
  volume = {91},
  pages = {035004},
  year = {2019},
  doi = {10.1103/RevModPhys.91.035004}
}

@article{Bernevig2005,
  author = {Bernevig, B. Andrei and Hughes, Taylor L. and Zhang, Shou-Cheng},
  title = {Orbitronics: The Intrinsic Orbital Current in \(p\)-Doped Silicon},
  journal = {Phys. Rev. Lett.},
  volume = {95},
  pages = {066601},
  year = {2005},
  doi = {10.1103/PhysRevLett.95.066601}
}

@article{Sohn2024,
  author  = {Sohn, Jeonghun and Lee, Jongjun M.
             and Lee, Hyun-Woo},
  title   = {{Dyakonov-Perel-like} Orbital and Spin Relaxations
             in Centrosymmetric Systems},
  journal = {Physical Review Letters},
  volume  = {132},
  number  = {24},
  pages   = {246301},
  year    = {2024},
  doi     = {10.1103/PhysRevLett.132.246301}
}

@article{Go2018,
  author = {Go, Dongwook and Jo, Daegeun and Kim, Changyoung and Lee, Hyun-Woo},
  title = {Intrinsic Spin and {Orbital Hall Effects} from Orbital Texture},
  journal = {Phys. Rev. Lett.},
  volume = {121},
  pages = {086602},
  year = {2018},
  doi = {10.1103/PhysRevLett.121.086602}
}

@article{Go2021EPL,
  author = {Go, Dongwook and Jo, Daegeun and Lee, Hyun-Woo and Kl{\"a}ui, Mathias and Mokrousov, Yuriy},
  title = {Orbitronics: Orbital currents in solids},
  journal = {EPL},
  volume = {135},
  pages = {37001},
  year = {2021},
  doi = {10.1209/0295-5075/ac2653}
}

@article{JoReview2024,
  author = {Jo, Daegeun and Go, Dongwook and Choi, Gyung-Min and Lee, Hyun-Woo},
  title = {Spintronics meets orbitronics: Emergence of orbital angular momentum in solids},
  journal = {npj Spintronics},
  volume = {2},
  pages = {19},
  year = {2024},
  doi = {10.1038/s44306-024-00023-6}
}

@article{AtenciaReview2024,
  author = {Burgos Atencia, Rhonald and Agarwal, Amit and Culcer, Dimitrie},
  title = {Orbital angular momentum of Bloch electrons: equilibrium formulation, magneto-electric phenomena, and the {orbital Hall effect}},
  journal = {Adv. Phys. X},
  volume = {9},
  pages = {2371972},
  year = {2024},
  doi = {10.1080/23746149.2024.2371972}
}

@article{Ando2025,
  author = {Ando, Kazuya},
  title = {Orbitronics: Harnessing orbital currents in solid-state devices},
  journal = {J. Phys. Soc. Jpn.},
  volume = {94},
  pages = {092001},
  year = {2025},
  doi = {10.7566/JPSJ.94.092001}
}

@article{Cysne2025,
  author = {Cysne, Tarik P. and Canonico, Luis M. and Costa, Marcio and Muniz, Roberto B. and Rappoport, Tatiana G.},
  title = {Orbitronics in two-dimensional materials},
  journal = {npj Spintronics},
  volume = {3},
  pages = {39},
  year = {2025},
  doi = {10.1038/s44306-025-00103-1}
}

@article{Lee2021,
  author = {Lee, Dongjoon and Go, Dongwook and Park, Hyeon-Jong and Jeong, Wonmin and Ko, Hye-Won and Yun, Dongwook and Jo, Daegeun and Lee, Soogil and Go, Gyungchoon and Oh, Ji-Hoon and Kim, Kab-Jin and Park, Byong-Guk and Min, Byoung-Chul and Koo, Hyun Cheol and Lee, Hyun-Woo and Lee, OukJae and Lee, Kyung-Jin},
  title = {Orbital torque in magnetic bilayers},
  journal = {Nat. Commun.},
  volume = {12},
  pages = {6710},
  year = {2021},
  doi = {10.1038/s41467-021-26650-9}
}

@article{Hayashi2024,
  author = {Hayashi, Hiroki and Go, Dongwook and Haku, Satoshi and Mokrousov, Yuriy and Ando, Kazuya},
  title = {Observation of orbital pumping},
  journal = {Nat. Electron.},
  volume = {7},
  pages = {646--652},
  year = {2024},
  doi = {10.1038/s41928-024-01193-1}
}

@article{Yang2024,
  author = {Yang, Yuhe and Wang, Ping and Chen, Jiali and Zhang, Delin and Pan, Chang and Hu, Shuai and Wang, Ting and Yue, Wensi and Chen, Cheng and Jiang, Wei and Zhu, Lujun and Qiu, Xuepeng and Yao, Yugui and Li, Yue and Wang, Wenhong and Jiang, Yong},
  title = {Orbital torque switching in perpendicularly magnetized materials},
  journal = {Nat. Commun.},
  volume = {15},
  pages = {8645},
  year = {2024},
  doi = {10.1038/s41467-024-52824-2}
}

@article{Wang2025,
  author = {Wang, Hanchen and Kang, Min-Gu and Petrosyan, Davit and Ding, Shilei and Schlitz, Richard and Riddiford, Lauren J. and Legrand, William and Gambardella, Pietro},
  title = {Orbital Pumping in Ferrimagnetic Insulators},
  journal = {Phys. Rev. Lett.},
  volume = {134},
  pages = {126701},
  year = {2025},
  doi = {10.1103/PhysRevLett.134.126701}
}

@article{Go2025OrbitalPumping,
  author  = {Go, Dongwook and Ando, Kazuya
             and Pezo, Armando and Bl{\"u}gel, Stefan
             and Manchon, Aur{\'e}lien and Mokrousov, Yuriy},
  title   = {Orbital pumping by magnetization dynamics
             in ferromagnets},
  journal = {Physical Review B},
  volume  = {111},
  number  = {14},
  pages   = {L140409},
  year    = {2025},
  doi     = {10.1103/PhysRevB.111.L140409}
}

@article{Keller2025,
  author = {Keller, Nils and Bose, Arnab and Soya, Nozomi and Hauth, Elias and Kammerbauer, Fabian and Gupta, Rahul and Hayashi, Hiroki and Kashiki, Hisanobu and Jakob, Gerhard and Krishnia, Sachin and Ando, Kazuya and Kl{\"a}ui, Mathias},
  title = {Identification of orbital pumping from spin pumping and rectification effects},
  journal = {Nano Lett.},
  volume = {25},
  pages = {13462--13467},
  year = {2025},
  doi = {10.1021/acs.nanolett.5c02641}
}

@article{Matsumoto2025,
  author = {Matsumoto, R. and Ohshima, R. and Ando, Y. and Go, Dongwook and Mokrousov, Yuriy and Shiraishi, M.},
  title = {Observation of Giant {Orbital Hall Effect} in {Si}},
  journal = {arXiv:2501.14237},
  year = {2025},
  eprint = {2501.14237},
  archivePrefix = {arXiv},
  doi = {10.48550/arXiv.2501.14237}
}

@book{Nye1985,
  author = {Nye, J. F.},
  title = {Physical Properties of Crystals: Their Representation by Tensors and Matrices},
  publisher = {Oxford University Press},
  address = {Oxford},
  year = {1985}
}

@article{Seemann2015,
  author = {Seemann, M. and K{\"o}dderitzsch, D. and Wimmer, S. and Ebert, H.},
  title = {Symmetry-imposed shape of linear response tensors},
  journal = {Phys. Rev. B},
  volume = {92},
  pages = {155138},
  year = {2015},
  doi = {10.1103/PhysRevB.92.155138}
}

@article{Luttinger1956,
  author = {Luttinger, J. M.},
  title = {Quantum Theory of Cyclotron Resonance in Semiconductors: General Theory},
  journal = {Phys. Rev.},
  volume = {102},
  pages = {1030--1041},
  year = {1956},
  doi = {10.1103/PhysRev.102.1030}
}

@article{Luttinger1955,
  author = {Luttinger, J. M. and Kohn, W.},
  title = {Motion of Electrons and Holes in Perturbed Periodic Fields},
  journal = {Phys. Rev.},
  volume = {97},
  pages = {869--883},
  year = {1955},
  doi = {10.1103/PhysRev.97.869}
}

@article{Kane1957,
  author = {Kane, E. O.},
  title = {Band structure of indium antimonide},
  journal = {J. Phys. Chem. Solids},
  volume = {1},
  pages = {249--261},
  year = {1957},
  doi = {10.1016/0022-3697(57)90013-6}
}

@book{Winkler2003,
  author = {Winkler, Roland},
  title = {Spin-Orbit Coupling Effects in Two-Dimensional Electron and Hole Systems},
  publisher = {Springer},
  address = {Berlin},
  year = {2003},
  doi = {10.1007/b13586}
}

@article{Culcer2004,
  author = {Culcer, D. and Sinova, J. and Sinitsyn, N. A. and Jungwirth, T. and MacDonald, A. H. and Niu, Q.},
  title = {Semiclassical Spin Transport in Spin-Orbit-Coupled Bands},
  journal = {Phys. Rev. Lett.},
  volume = {93},
  pages = {046602},
  year = {2004},
  doi = {10.1103/PhysRevLett.93.046602}
}

@article{Kane_2,
  author = {Kane, E. O.},
  title = {The \({\bf k}\cdot {\bf p}\) method},
  journal = {Semicond. Semimetals},
  volume = {1},
  pages = {75--100},
  year = {1966}
}

@book{supplement,
  note={Refer to the Supplemental Material at [...] for the derivation of the renormalized orbital angular-momentum density, current, and torque operators; the OAM matrices and spin-orbital textures of group-IV semiconductors; the formulation of the quantum scattering theory of spin pumping;  and the symmetry analysis of spin-orbital conversion in different materials, together with additional results for Si}
}

@incollection{Brataas2012,
  author = {Brataas, Arne and Tserkovnyak, Yaroslav and Bauer, Gerrit E. W. and Kelly, Paul J.},
  title = {Spin Pumping and Spin Transfer},
  booktitle = {Spin Current},
  editor = {Maekawa, S. and Valenzuela, S. O. and Saitoh, E. and Kimura, T.},
  publisher = {Oxford University Press},
  address = {Oxford},
  pages = {87--135},
  year = {2012},
  doi = {10.1093/acprof:oso/9780199600380.003.0008}
}

@article{Bahder1990,
  author  = {Bahder, Thomas B.},
  title   = {Eight-band {$\mathbf{k}\cdot\mathbf{p}$} model of strained zinc-blende crystals},
  journal = {Phys. Rev. B},
  volume  = {41},
  number  = {17},
  pages   = {11992--12001},
  year    = {1990},
  doi = {10.1103/PhysRevB.41.11992}
}

@article{Lawaetz1971,
  author  = {Lawaetz, P.},
  title   = {Valence-Band Parameters in Cubic Semiconductors},
  journal = {Phys. Rev. B},
  volume  = {4},
  number  = {10},
  pages   = {3460--3467},
  year    = {1971},
  doi = {10.1103/PhysRevB.4.3460}
}

@article{Orbital_Pumping2026,
  author  = {Cai, C. and Wang, H. and Yu, T.},
  title   = {Evanescent orbital pumping by magnetization dynamics free of spin-orbit coupling},
  journal = {Sci. China Phys. Mech. Astron.},
  volume  = {69},
  pages   = {237511},
  year    = {2026},
  doi = {10.1007/s11433-025-2833-1}
}

@article{kramers,
  title={Th{\'e}orie g{\'e}n{\'e}rale de la rotation paramagn{\'e}tique dans les cristaux},
  author={Kramers, Hendrik Anthony},
  journal={Proc. Roy. Acad. Amsterdam},
  volume={33},
  pages={959--972},
  year={1930}
}

@article{Chen2015SpinPumpingSOC,
  author  = {Chen, Kai and Zhang, Shufeng},
  title   = {Spin Pumping in the Presence of Spin-Orbit Coupling},
  journal = {Phys. Rev. Lett.},
  volume  = {114},
  pages   = {126602},
  year    = {2015},
  doi = {10.1103/PhysRevLett.114.126602}
}

@article{Santos2024GeOHE,
  author = {Santos, E. and Abr{\~a}o, J. E. and Costa, J. L.
            and Santos, J. G. S. and Rodrigues-Junior, G.
            and Mendes, J. B. S. and Azevedo, A.},
  title = {Negative {orbital Hall effect} in germanium},
  journal = {Physical Review Applied},
  volume = {22},
  pages = {064071},
  year = {2024},
  doi = {10.1103/PhysRevApplied.22.064071}
}

@article{RojasSanchez2013,
  author = {Rojas-S{\'a}nchez, J.-C. and Cubukcu, M.
            and Jain, A. and Vergnaud, C.
            and Portemont, C. and Ducruet, C.
            and Barski, A. and Marty, A.
            and Vila, L. and Attan{\'e}, J.-P.
            and Augendre, E. and Desfonds, G.
            and Gambarelli, S. and Jaffr{\`e}s, H.
            and George, J.-M. and Jamet, M.},
  title = {Spin pumping and inverse {spin Hall effect} in germanium},
  journal = {Physical Review B},
  volume = {88},
  pages = {064403},
  year = {2013},
  doi = {10.1103/PhysRevB.88.064403}
}

@article{Kajiwara2010,
  author = {Kajiwara, Y. and Harii, K. and Takahashi, S.
            and Ohe, J. and Uchida, K. and Mizuguchi, M.
            and Umezawa, H. and Kawai, H. and Ando, K.
            and Takanashi, K. and Maekawa, S. and Saitoh, E.},
  title = {Transmission of electrical signals by spin-wave
           interconversion in a magnetic insulator},
  journal = {Nature},
  volume = {464},
  pages = {262--266},
  year = {2010},
  doi = {10.1038/nature08876}
}

@article{Sahu2024,
  author  = {Sahu, Pratik and Bidika, Jatin Kumar and Biswal, Bubunu and
             Satpathy, S. and Nanda, B. R. K.},
  title   = {Emergence of giant orbital Hall and tunable {spin Hall effects}
             in centrosymmetric transition metal dichalcogenides},
  journal = {Phys. Rev. B},
  volume  = {110},
  number  = {5},
  pages   = {054403},
  year    = {2024},
  doi     = {10.1103/PhysRevB.110.054403}
}

@article{Sahu2021,
  author  = {Sahu, Pratik and Bhowal, Sayantika and Satpathy, S.},
  title   = {Effect of the inversion symmetry breaking on the {orbital
             Hall effect}: A model study},
  journal = {Phys. Rev. B},
  volume  = {103},
  number  = {8},
  pages   = {085113},
  year    = {2021},
  doi     = {10.1103/PhysRevB.103.085113}
}

@article{Canonico2020,
  author  = {Canonico, Luis M. and Cysne, Tarik P. and
             Molina-Sanchez, Alejandro and Muniz, R. B. and
             Rappoport, Tatiana G.},
  title   = {Orbital Hall insulating phase in transition metal
             dichalcogenide monolayers},
  journal = {Phys. Rev. B},
  volume  = {101},
  number  = {16},
  pages   = {161409},
  year    = {2020},
  doi     = {10.1103/PhysRevB.101.161409}
}

@article{Cysne2021,
  author  = {Cysne, Tarik P. and Costa, Marcio and Canonico, Luis M. and
             Nardelli, M. Buongiorno and Muniz, R. B. and
             Rappoport, Tatiana G.},
  title   = {Disentangling orbital and valley Hall effects in bilayers
             of transition metal dichalcogenides},
  journal = {Phys. Rev. Lett.},
  volume  = {126},
  number  = {5},
  pages   = {056601},
  year    = {2021},
  doi     = {10.1103/PhysRevLett.126.056601}
}

@article{Costa2023,
  author  = {Costa, Marcio and Focassio, Bruno and Canonico, Luis M. and
             Cysne, Tarik P. and Schleder, Gabriel R. and Muniz, R. B. and
             Fazzio, Adalberto and Rappoport, Tatiana G.},
  title   = {Connecting higher-order topology with the orbital Hall
             effect in monolayers of transition metal dichalcogenides},
  journal = {Phys. Rev. Lett.},
  volume  = {130},
  number  = {11},
  pages   = {116204},
  year    = {2023},
  doi     = {10.1103/PhysRevLett.130.116204}
}

@article{Cysne2023,
  author  = {Cysne, Tarik P. and Guimaraes, Filipe S. M. and
             Canonico, Luis M. and Costa, Marcio and
             Rappoport, Tatiana G. and Muniz, R. B.},
  title   = {Orbital magnetoelectric effect in nanoribbons of
             transition metal dichalcogenides},
  journal = {Phys. Rev. B},
  volume  = {107},
  number  = {11},
  pages   = {115402},
  year    = {2023},
  doi     = {10.1103/PhysRevB.107.115402}
}

@article{Fukami2026,
  author  = {Fukami, Shunsuke and Lee, Kyung-Jin and
             Kl{\"a}ui, Mathias},
  title   = {Challenges and opportunities in orbitronics},
  journal = {Nature Physics},
  volume  = {22},
  pages   = {999--1005},
  year    = {2026},
  doi     = {10.1038/s41567-025-03143-w}
}

@article{Pezo2025,
  author  = {Pezo, Armando and Go, Dongwook and
             Mokrousov, Yuriy and Jaffr{\`e}s, Henri and
             Manchon, Aur{\'e}lien},
  title   = {Adiabatic spin and orbital pumping in
             metallic heterostructures},
  journal = {Phys. Rev. B},
  volume  = {111},
  number  = {13},
  pages   = {134424},
  year    = {2025},
  doi     = {10.1103/PhysRevB.111.134424}
}

@article{Han2025,
  author  = {Han, Seungyun and Ko, Hye-Won and
             Oh, Jung Hyun and Lee, Hyun-Woo and
             Lee, Kyung-Jin and Kim, Kyoung-Whan},
  title   = {Orbital Pumping Incorporating Both
             Orbital Angular Momentum and Position},
  journal = {Phys. Rev. Lett.},
  volume  = {134},
  number  = {3},
  pages   = {036305},
  year    = {2025},
  doi     = {10.1103/PhysRevLett.134.036305}
}

@article{Ning2026,
  author  = {Ning, Xiaobai and Jaffr{\`e}s, Henri and
             Zhao, Weisheng and Manchon, Aur{\'e}lien},
  title   = {Phenomenology of orbital torque, pumping,
             and mixing conductance in metallic bilayers},
  journal = {Phys. Rev. Research},
  volume  = {8},
  number  = {2},
  pages   = {023239},
  year    = {2026},
  doi     = {10.1103/h3tq-rrgx}
}

@article{Huang2018,
  author  = {Huang, Chunli and Tokatly, Ilya V. and
             Bergeret, F. Sebastian},
  title   = {Extrinsic spin-charge coupling in diffusive
             superconducting systems},
  journal = {Phys. Rev. B},
  volume  = {98},
  number  = {14},
  pages   = {144515},
  year    = {2018},
  doi     = {10.1103/PhysRevB.98.144515}
}

@article{Kokkeler2025,
  author  = {Kokkeler, Tim and Bergeret, F. Sebastian and
             Tokatly, I. V.},
  title   = {Universal Phenomenology of Charge-Spin
             Interconversion and Dynamics in Diffusive
             Systems with Spin-Orbit Coupling},
  journal = {Phys. Rev. Lett.},
  volume  = {134},
  number  = {9},
  pages   = {096001},
  year    = {2025},
  doi     = {10.1103/PhysRevLett.134.096001}
}

@article{RyooPark2017,
  author  = {Ryoo, Ji Hoon and Park, Cheol-Hwan},
  title   = {Hidden orbital polarization in diamond, silicon, germanium, gallium arsenide and layered materials},
  journal = {NPG Asia Materials},
  volume  = {9},
  pages   = {e382},
  year    = {2017},
  doi     = {10.1038/am.2017.67}
}

@article{Zhang2014,
  author  = {Zhang, Xiuwen and Liu, Qihang and Luo, Jun-Wei and Freeman, Arthur J. and Zunger, Alex},
  title   = {Hidden spin polarization in inversion-symmetric bulk crystals},
  journal = {Nature Physics},
  volume  = {10},
  number  = {5},
  pages   = {387--393},
  year    = {2014},
  doi     = {10.1038/nphys2933}
}

@article{RangKelly2024,
  author  = {Rang, Max and Kelly, Paul J.},
  title   = {Orbital relaxation length from first-principles
             scattering calculations},
  journal = {Physical Review B},
  volume  = {109},
  number  = {21},
  pages   = {214427},
  year    = {2024},
  doi     = {10.1103/PhysRevB.109.214427}
}

@article{ValetPRB2025,
  author  = {Valet, Thierry and Raimondi, Roberto},
  title   = {Quantum kinetic theory of the linear response
             for weakly disordered multiband systems},
  journal = {Physical Review B},
  volume  = {111},
  pages   = {L041118},
  year    = {2025},
  doi     = {10.1103/PhysRevB.111.L041118}
}

@article{ValetPRL2025,
  author  = {Valet, Thierry and Jaffres, Henri and
             Cros, Vincent and Raimondi, Roberto},
  title   = {Quantum Kinetic Anatomy of Electron Angular
             Momenta Edge Accumulation},
  journal = {Physical Review Letters},
  volume  = {135},
  number  = {25},
  pages   = {256301},
  year    = {2025},
  doi     = {10.1103/86cm-yn9z}
}

\section{End Matter}

\textit{Relation to conventional spin pumping}.---The general formula \eqref{pumped_current} is reduced to the longitudinal spin current when considering a free electron model with dispersion $\varepsilon^h_{n}({\bf k})=\hbar^2k^2/(2m^*)$ and effective mass $m^*$~\cite{Tserkovnyak2002PRL,Tserkovnyak2005RMP}, in which the band indices are reduced to the spin indices $\{m,n,m'\}\in\{\uparrow,\downarrow\}$.
The summation over the spin indices $\{m,n,m'\}$ then simplifies the pumped spin current to 
\[
{J}^{{\rm DC},i}_{S_\alpha}\propto\int d{\pmb \kappa}dq_z F(\pmb{\kappa},q_z)({\bf M}^{(+)}\times{\bf M}^{(+)*})_\alpha{\cal V}_i(-\tilde{q}_z,-\pmb{\kappa}),
\]
where $F({\pmb \kappa},q_z)$ is an even function of wave vectors. The spin polarization of the pumped current is aligned with the saturation magnetization, along the direction of $\mathbf{M}^{(+)}\times\mathbf{M}^{(+)*}$. Further, since ${\cal V}_{x,y}(-\tilde{q}_z,-\pmb{\kappa})$ are odd under $\pmb{\kappa}\rightarrow-\pmb{\kappa}$, only the longitudinal spin current flowing normal to the interface survives upon integration over the in-plane momenta~\cite{Tserkovnyak2002PRL,Tserkovnyak2005RMP}. This process can be understood in terms of spin-dependent scattering, whereby holes are transferred from states with spin antiparallel to the saturation magnetization to states with spin parallel to it.

By contrast, in multiorbital materials with spin-orbital textures,  the summation over band indices becomes considerably more intricate and generally cannot be simplified in the same manner. As a consequence, the spin polarization of the pumped current is no longer necessarily aligned with the saturation magnetization, and spin current flowing in the other directions may emerge.

\textit{Renormalized orbital matrix by SOC}.---As addressed in the main text, the SOC with the bands out of the subspace of interest efficiently renormalizes the orbital matrix of holes, which in the orbital-spin product basis $|P_{x,y,z}\rangle\otimes|s=\{\uparrow,\downarrow\}\rangle$ is found to be (see SM~\cite{supplement})
\begin{widetext}
\begin{align}
    &{\cal L}_x =\frac{m_0}{\hbar^2}\left(
\begin{array}{cccccc}
0 & 0 & 0 & \frac{i}{3}(D_1 - D_2) & -\frac{1}{3}(D_1 - D_2) & 0 \\
0 & 0 & -\frac{i}{3}(D_1 - D_2) & 0 & 0 & \frac{1}{3}(D_1 - D_2) \\
0 & \frac{i}{3}(D_1 - D_2) & 0 & \frac{2}{3}(D_1 - D_2) & -\frac{2}{3}i(2D_1 + D_2) & 0 \\
-\frac{i}{3}(D_1 - D_2) & 0 & \frac{2}{3}(D_1 - D_2) & 0 & 0 & -\frac{2}{3}i(2D_1 + D_2) \\
-\frac{1}{3}(D_1 - D_2) & 0 & \frac{2}{3}i(2D_1 + D_2) & 0 & 0 & \frac{2}{3}(D_1 - D_2) \\
0 & \frac{1}{3}(D_1 - D_2) & 0 & \frac{2}{3}i(2D_1 + D_2) & \frac{2}{3}(D_1 - D_2) & 0
\end{array}
\right),\nonumber\\
&{\cal L}_y =\frac{m_0}{\hbar^2}\left(
\begin{array}{cccccc}
0 & -\frac{2}{3}i(D_1 - D_2) & 0 & -\frac{1}{3}(D_1 - D_2) & \frac{2}{3}i(2D_1 + D_2) & 0 \\
\frac{2}{3}i(D_1 - D_2) & 0 & -\frac{1}{3}(D_1 - D_2) & 0 & 0 & \frac{2}{3}i(2D_1 + D_2) \\
0 & \frac{1}{3}(-D_1 + D_2) & 0 & 0 & -\frac{1}{3}(D_1 - D_2) & 0 \\
-\frac{1}{3}(D_1 - D_2) & 0 & 0 & 0 & 0 & \frac{1}{3}(D_1 - D_2) \\
-\frac{2}{3}i(2D_1 + D_2) & 0 & -\frac{1}{3}(D_1 - D_2) & 0 & 0 & -\frac{2}{3}i(D_1 - D_2) \\
0 & -\frac{2}{3}i(2D_1 + D_2) & 0 & \frac{1}{3}(D_1 - D_2) & \frac{2}{3}i(D_1 - D_2) & 0
\end{array}
\right),\nonumber\\
&{\cal L}_z =\frac{m_0}{\hbar^2}\left(
\begin{array}{cccccc}
\frac{2}{3}(D_1 - D_2) & 0 & -\frac{2}{3}i(2D_1 + D_2) & 0 & 0 & -\frac{1}{3}(D_1 - D_2) \\
0 & -\frac{2}{3}(D_1 - D_2) & 0 & -\frac{2}{3}i(2D_1 + D_2) & -\frac{1}{3}(D_1 - D_2) & 0 \\
\frac{2}{3}i(2D_1 + D_2) & 0 & \frac{2}{3}(D_1 - D_2) & 0 & 0 & \frac{1}{3}i(D_1 - D_2) \\
0 & \frac{2}{3}i(2D_1 + D_2) & 0 & -\frac{2}{3}(D_1 - D_2) & -\frac{1}{3}i(D_1 - D_2) & 0 \\
0 & -\frac{1}{3}(D_1 - D_2) & 0 & \frac{1}{3}i(D_1 - D_2) & 0 & 0 \\
-\frac{1}{3}(D_1 - D_2) & 0 & -\frac{1}{3}i(D_1 - D_2) & 0 & 0 & 0
\end{array}
\right).
\end{align}
\end{widetext}

Particularly, disregarding the SOC, i.e.  $\lambda=0$, renders the coefficients $D_1=D_2$, leading to ${\pmb{\cal L}}=(m_0D/\hbar^2){\pmb{\cal I}}$ with $D=D_1+D_2$, where ${\pmb{\cal I}}$ denotes the atomic orbital angular momentum~\cite{supplement}. This approximation was used to obtain orbital dynamics in semiconductors~\cite{Orbital_Pumping2026,Bernevig2005,Luttinger1956}.

For the group-IV semiconductors with the diamond structure, such as Si, Ge, Sn, and diamond, we use parameters in the multiband Kane Hamiltonian~\cite{Kane1957,Kane_2,Winkler2003,Bahder1990} to calculate the orbital matrix, as well as its associated orbital density and orbital-current density operators of the valence bands, as listed in Table~\ref{tab3}~\cite{Lawaetz1971}.
The constants $m_0D_1/\hbar^2=-7.39$ and $m_0D_2/\hbar^2=-5.57$ differ for Ge, but they are almost the same for Si and diamond. Intriguingly, $m_0D_1/\hbar^2=14.41$ and $m_0D_2/\hbar^2=-16.67$ can have different signs for Sn. The difference between $D_1$ and $D_2$ arises from the atomic SOC, which modifies the orbital angular momentum of $p$-orbital electrons and causes it to deviate from that of the atomic one. The summation $m_0(D_1+D_2)/\hbar^2=-12.96$ for Ge is approximately equal to the result $m_0D/\hbar^2=-11.2$, given in Refs~\cite{Luttinger1955,Luttinger1956}.

\begin{table}[h]
\caption{\label{tab3} Kane-model parameters and orbital parameters for group-IV elements~\cite{Lawaetz1971}.}
\begin{tabular}{lcccccc}
\toprule
Element & $E_0$ (eV) & $\lambda$ (eV) & $P$ (\AA$\cdot$eV) & ${m_0 D_1}/{\hbar^2}$ & ${m_0 D_2}/{\hbar^2}$ \\
\midrule
Si & $4.07$  & $0.04$ & $9.07$ & $-1.33$ & $-1.31$ \\
Ge & $0.89$  & $0.29$ & $10.01$  & $-7.39$ & $-5.57$ \\
Sn & $-0.41$  & $0.77$ & $9.52$  & $14.41$   & $-16.67$ \\
C  & $13.04$ & $0.00$ & $13.78$ & $-0.95$ & $-0.95$ \\
\bottomrule
\end{tabular}
\end{table}

\textit{Symmetry analysis and wide material choices}.--- Due to the SOC, the pumped longitudinal spin current is converted into a longitudinal orbital current flowing perpendicular to the interface, as well as an orbital Hall current propagating parallel to the interface. The conversion from the orbital current in turn induces the transverse components of the spin current. The conversion from the spin current ${\bf J}_S$ to orbital current ${\bf J}_L$ is expressed as
\begin{align}
J_{L,\alpha}^i=N_{\alpha\beta}^{ij}J_{S,\beta}^j,
\end{align}
governed by spin-orbital conversion matrix $N_{\alpha \beta}^{ij}$. Because the spin current $J_{S, \beta}^{j}$ and orbital current $J_{L, \alpha}^{i}$ are second-order axial tensors, in which the subscript indicates the angular-momentum polarization and the superscript represents the spatial indices, the coefficients $[N_{\alpha \beta}^{i j}]$ form a fourth-order polar tensor with $3\times3\times3\times3=81$ elements. It is constrained by the crystal point-group symmetry
\begin{align}
N_{\alpha \beta}^{i j}=\sum_{m, n, \eta, \lambda=\{x,y,z\}} R_{i m} R_{j n} R_{\alpha \eta} R_{\beta \lambda} N_{\eta \lambda}^{m n},
\end{align}
where $R$ is the transformation matrix of the symmetry operation.

The group-IV semiconductors such as Ge, Si, and diamond exhibit the $m\overline{3}m~(O_h)$ point-group symmetry. The symmetry analysis implies that the nonzero orbital-spin conversion tensor $[N_{\alpha \beta}^{i j}]$ has four independent coefficients $\{N_1,N_2,N_3,N_4\}$ as shown in  Table~\ref{interconversion}. 
When the saturation magnetization is perpendicular to the interface as in Fig.~\ref{swapping_current}(a), the primary pumped spin current $J_{S_z}^z$ is converted into orbital currents $J_{L_x}^x=J_{L_y}^y=N_2J_{S_z}^z$ and $J_{L_z}^z=N_1J_{S_z}^z$. In contrast, when the saturation magnetization along $\hat{\bf y}$ is parallel to the interface as in Fig.~\ref{swapping_current}(c), the primary spin current $J_{S_y}^z$ can generate a longitudinal orbital current $J_{L_y}^z=N_3J_{S_y}^z$ and a swapped orbital current $J_{L_z}^y=N_4J_{S_y}^z$. The same symmetry analysis applies to the conversion from the orbital current to the swapped spin current. Our numerical results for Ge in Fig.~\ref{swapping_current} are consistent with these symmetry-allowed current components and their relations. Furthermore, extending the symmetry analysis beyond group-IV semiconductors to materials with different crystal symmetries, \textit{e.g.,} GaAs, GaN, CuInSe$_2$, and SnSe, reveals that analogous swapping channels are also symmetry-allowed (see SM~\cite{supplement}). These findings suggest that spin and orbital swapping in the pumping configuration can occur in a broad range of multiorbital materials.

\begin{table}[h]
\centering
\caption{Spin-orbital interconversion tensor $[N_{\alpha \beta}^{i j}]$.}
\label{interconversion}
\renewcommand{\arraystretch}{1.2} 
\setlength{\tabcolsep}{3.5pt} 
\begin{tabular}{c|ccccccccc} 
\hline
$i\alpha\backslash j\beta$ & $xx$ & ~~$yx$ & ~~$zx$ & ~~$xy$ & ~~$yy$ & ~~$zy$ & ~~$xz$ & ~~$yz$ & ~~$zz$ \\
\hline
$xx$ & $N_1$ & 0 & 0 & 0 & $N_2$ & 0 & 0 & 0 & $N_2$ \\
$yx$ & 0 & $N_3$ & 0 & $N_4$ & 0 & 0 & 0 & 0 & 0 \\
$zx$ & 0 & 0 & $N_3$ & 0 & 0 & 0 & $N_4$ & 0 & 0 \\
$xy$ & 0 & $N_4$ & 0 & $N_3$ & 0 & 0 & 0 & 0 & 0 \\
$yy$ & $N_2$ & 0 & 0 & 0 & $N_1$ & 0 & 0 & 0 & $N_2$ \\
$zy$ & 0 & 0 & 0 & 0 & 0 & $N_3$ & 0 & $N_4$ & 0 \\
$xz$ & 0 & 0 & $N_4$ & 0 & 0 & 0 & $N_3$ & 0 & 0 \\
$yz$ & 0 & 0 & 0 & 0 & 0 & $N_4$ & 0 & $N_3$ & 0 \\
$zz$ & $N_2$ & 0 & 0 & 0 & $N_2$ & 0 & 0 & 0 & $N_1$ \\
\hline
\end{tabular}
\end{table}

\end{document}